# Purcell-enhanced emissions from diamond color centers in slow light photonic crystal waveguides

Sophie W. Ding[1,#,†], Chang Jin[1,#] Kazuhiro Kuruma[1,2,#,*], Xinghan Guo[3], Michael Haas[1], Boris Korzh[4], Andrew Beyer[4], Matt Shaw[4], Neil Sinclair[1], David D. Awschalom[3,5], F. Joseph Heremans[3,5], Nazar Delegan[5], Alexander A. High[3,5]*, and Marko Loncar[1,*]

[1]John A. Paulson School of Engineering and Applied Sciences, Harvard University, Cambridge, Massachusetts, USA.
[2]Research Center for Advanced Science and Technology, The University of Tokyo, Meguro-ku, Tokyo, Japan.
[3]Pritzker School of Molecular Engineering, University of Chicago, Chicago, IL, USA.
[4]Jet Propulsion Laboratory, California Institute of Technology, 4800 Oak Grove Dr., Pasadena, California 91109, USA
[5]Q-NEXT, Argonne National Laboratory, Lemont, IL.
*E-mail: kkuruma@g.ecc.u-tokyo.ac.jp; ahigh@uchicago.edu; loncar@g.harvard.edu
#These authors contributed equally: Sophie W. Ding, Chang Jin, Kazuhiro Kuruma.
†Present address: AWS Center for Quantum Computing, San Francisco, California, USA.

**Abstract**

Quantum memories based on emitters with optically addressable spins rely on efficient photonic interfaces, often implemented as nanophotonic cavities with ideally narrow spectral linewidths and small mode volumes. However, these approaches require nearly perfect spectral and spatial overlap between the cavity mode and quantum emitter, which can be challenging. This is especially true in the case of solid-state quantum emitters that are often randomly positioned and can suffer from significant inhomogeneous broadening. An alternative approach to mitigate these challenges is to use slow-light waveguides that can enhance light-matter interaction across large optical bandwidths and large areas. Here, we demonstrate diamond slow light photonic crystal (PhC) waveguides that enable broadband optical coupling to embedded silicon-vacancy (SiV) color centers. We take advantage of the recently demonstrated thin-film diamond photonic platform to fabricate fully suspended two-dimensional PhC waveguides. Using this approach, we demonstrate waveguide modes with high group indices up to 70 and observe Purcell-enhanced emissions of the SiVs coupled to the waveguide mode. Our approach represents a practical diamond platform for robust spin-photon interfaces with color centers.

**Introduction**

Color centers in diamond have emerged as promising platforms for quantum information processing, and have been used as single photon sources[1–3], as well as quantum memories by leveraging their long-lived and optically accessible spins[45,64]. Additionally, they play an important role in scalable quantum networks, where long-distance entanglement between multiple color centers has been demonstrated[7–9]. For practical quantum applications, achieving high emission rates and collection efficiencies of emitted photons are key. As a result, a wide variety of diamond nanophotonic structures have been explored, including optical cavities[10–13]. Among them, diamond PhC cavities, with their small mode volumes (V) and relatively high quality factors (Q)[14–16], are considered the most promising and have been the workhorses behind various quantum network demonstrations[5,6].

However, large Q and small V put stringent constraints on the spectral and spatial overlap between the color center and the cavity mode, which can be challenging to achieve. For example, fabrication tolerances can significantly affect the resonant wavelength of the fabricated cavities, while variations in the local environment of each emitter (e.g. strain, charge, etc) can result in inhomogeneous broadening[17,18]. While spatial mismatch can be addressed using techniques based on deterministic positioning of color centers in nanostructures[19,20], the spectral mismatch requires tuning either the cavity resonance, e.g. using gas condensation[21,22] or surface oxidation/deposition[11,14,23], or the emitter, e.g by strain[24,25]. These, however, can be challenging to implement in a scalable manner [26,27].

An alternative approach to achieve an efficient photonic interface is to use slow light waveguides. Slow light PhC waveguides can enable broadband optical coupling to solid-state emitters as well as enhance spontaneous emission of the emitters by the Purcell effect in the slow light region[28–31], leading to near-unity cavity-emitter coupling efficiency ($\beta$)[32]. PhC waveguides can also allow for chiral coupling to quantum emitters with circularly polarized emissions[33], as well as enable efficient topologically-protected nanophotonic devices[34,35]. Two-dimensional (2D) PhC waveguides could also offer better thermalization and optical power handling[36], which are critical for the required cryogenic temperature for quantum operation. Despite these advantages, 2D slow light PhC waveguides in diamond or their coupling to color centers have not been demonstrated. This is likely due to the fabrication challenges associated with fabricating suspended 2D diamond devices compatible with color centers.

Here, we demonstrate 2D slow light PhC waveguides coupled to silicon-vacancy centers (SiVs) in diamond. We fabricate the PhC waveguides on a single crystalline thin film diamond created by scalable fabrication techniques with a combination of ion implantation, diamond overgrowth, and thin film transfer-printing processes[37,38]. We show high group indices ($n_g$) up to 70 in the slow-light regime through photoluminescence measurements of the waveguide modes. We also observe Purcell-enhanced emission from (single) SiVs coupled to the PhC waveguide mode, with a ~2-fold increase in the spontaneous emission rate in the slow-light regime compared to those in the slab. These findings present a scalable pathway for efficient quantum photonic platforms with diamond color centers.

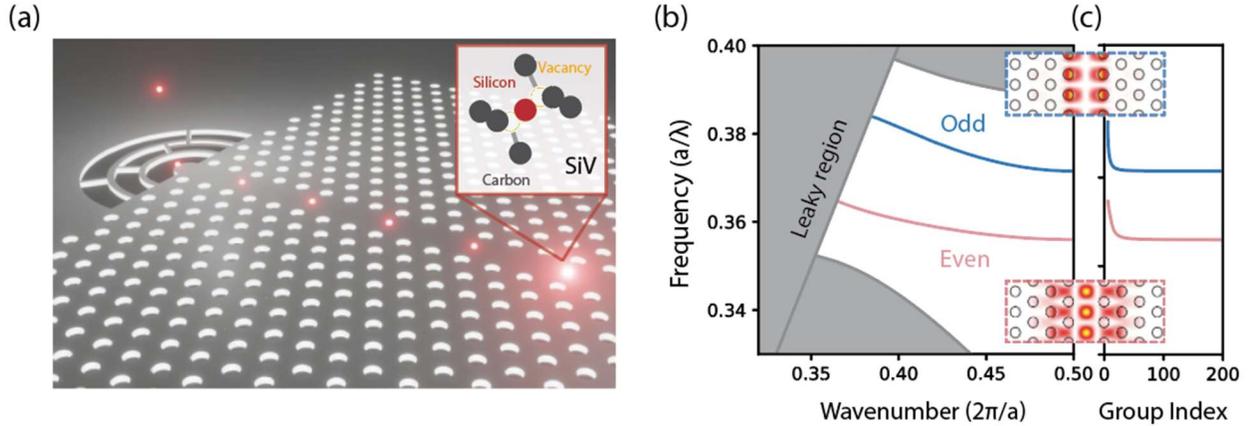

**Fig. 1.** (a) A schematic view of a suspended PhC diamond waveguide with embedded (single) SiVs. Input/output grating couplers are located at both ends of the waveguide (only one side is shown in the figure). The red dots signify the photons emitted by the SiV guided by the waveguide and coupled out through the grating coupler. (b) Simulated dispersion relation for the slow-light waveguide. Red and blue curves are even (0th-order) and odd (1st-order) modes, respectively. *a* is the lattice constant of the photonic crystal. (c) Corresponding group index of the two modes. The insets show the simulated electric field ($|E_y|^2$) profiles of the even and odd modes with $n_g$ of ~20 at 0.357($a/\lambda$) and 0.373($a/\lambda$), respectively.

As schematically shown in Fig. 1(a), the diamond PhC waveguide is formed by removing one row in the 2D PhCs. We vary the lattice constants in fabrication, where $a$=255-275 nm and an air hole radius (*r*) of 65 nm. The thickness of the diamond slab (*d*) is 160 nm. Fig. 1(b) shows the band diagram for the PhC waveguide calculated using the 3D plane wave expansion (PWE) method. The refractive index of the diamond slab is assumed to be $n$ = 2.4. Two distinct guided waveguide modes are found inside the photonic bandgap, spanning the frequency range $a/\lambda$ =0.353-0.389, where $\lambda$ is the wavelength. The lower frequency waveguide mode corresponds to the even (0th-order) mode (lower inset), while the higher frequency mode is the odd (1st-order) waveguide mode (upper inset). The band structure of these waveguide modes is consistent with previous reports using PhC waveguides[30,39]. Fig. 1(c) shows the calculated group index ($n_g$), obtained by $n_g = c(dk/d\omega)$, using the data from Fig. 1(b). For both modes, group indices exceeding 100 can be found near the edge of Brillouin zone. In our experiments, we focus on the even waveguide mode to demonstrate coupling with SiVs.

We fabricate the designed waveguide on a 160 nm-thick diamond membrane with a surface roughness < 0.3 nm and thickness variation of ~ 1 nm. The detailed fabrication and properties of the diamond membrane can be found in previous work[37,38]. Fig. 2(a) shows the simplified flow of the fabrication process[16]. The diamond membrane (200 μm × 200 μm) is implanted with Si ions resulting in randomly positioned SiVs. The implantation depth is 40 nm below the diamond surface. The membrane is transferred onto a $SiO_2$-on-Si substrate (10 mm × 10 mm) by a transfer-printing technique for handling (step 1). Next, we deposit Au/Cr at the edges of the membrane via a liftoff process to secure the membrane and minimize the chances of delamination during the fabrication process (step 2). We then deposit 100 nm-thick SiN and 400 nm-thick electron beam (EB) resist (ZEP-520A) on the membrane (step 3). EB lithography is used to write the waveguide structure, which is then transferred into the SiN mask layer by inductively coupled plasma reactive ion etching (ICP-RIE) using $SF_6$ and $C_4F_8$ gases (step 4). We then transfer the waveguide pattern from the mask into the diamond membrane using $O_2$-plasma RIE (step 5). The SiN and $SiO_2$ layers are removed by immersion in hydrofluoric acid (HF). Finally, we perform additional undercutting using $XeF_2$ gas to etch the Si substrate underneath to maintain enough air gap between the diamond slab and Si substrate (step 6).

An optical microscope image of the diamond membrane with fabricated arrays of PhC waveguides is shown in Fig. 2(b). Fig. 2(c) shows the top view of a scanning electron microscope (SEM) image of a fabricated PhC waveguide. The position of grating couplers is slightly shifted from the center of the waveguide to increase the visibility of Fabry-Perot (FP) fringes[34] (Fig. 2(c)). Fig. 2(d) and (e) display the zoom-in view of the PhC waveguide and grating coupler region, respectively. We emphasize that our fabrication method does not involve complicated bulk diamond undercut processes that can result in additional surface roughness[40,41]. This is critical for realizing high-quality photonic devices in diamond[16], especially those based on 2D photonic crystal structures.

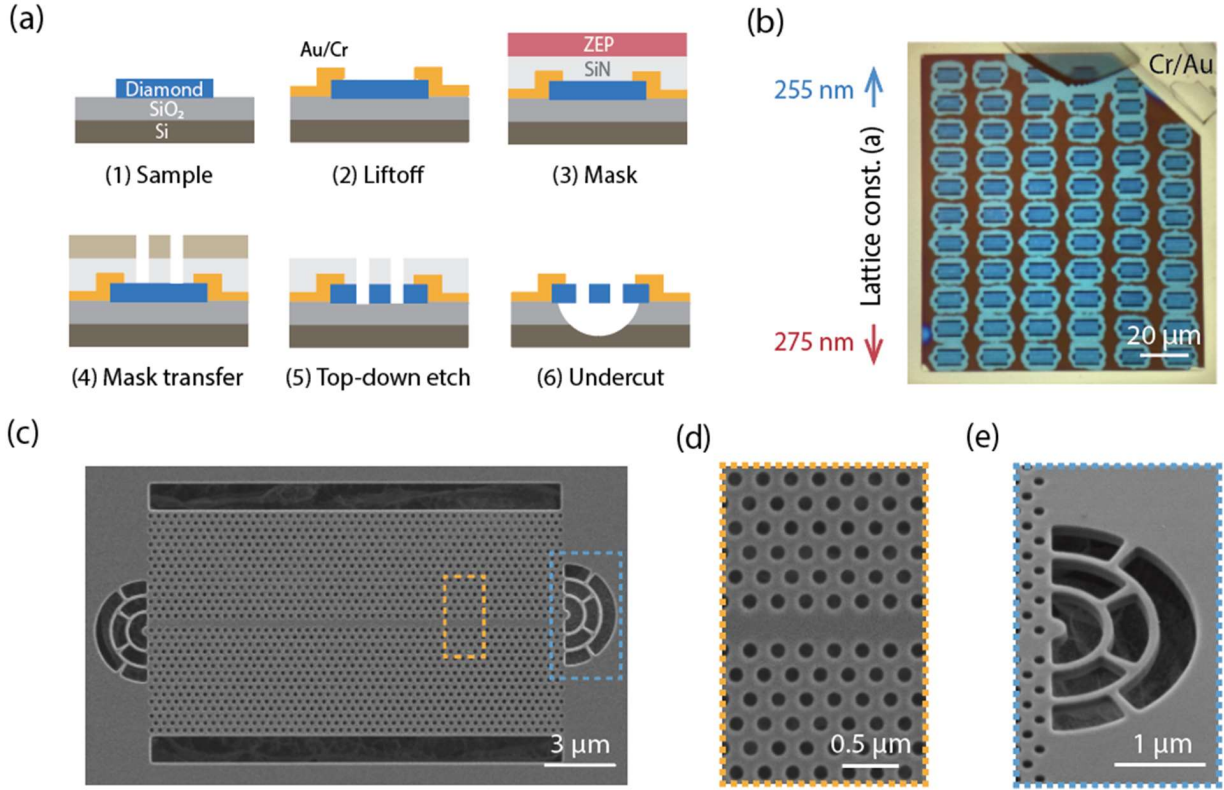

**Fig. 2.** (a) Process flow for fabrication of PhC waveguides in a thin film diamond: (1) thin film diamond is transferred onto SiO2/Si substrate; (2) Au/Cu pad is created around the diamond by a liftoff process; (3) EB resist (ZEP-520A) and SiN layer are formed on the diamond; (4) the pattern on the EB resist is transferred unto the SiN mask; (5) the waveguide structure is patterned into the diamond by ICP-RIE-based top-down etching; (6) The undercut process is performed by etching SiO2 and Si underneath. (b) Optical microscope image of a fabricated PhC waveguide array in a thin film diamond (200 × 200 μm) on a Si substrate. The thin film is secured on the Si substrate with Au/Cr pads. (c) SEM images of one of the fabricated PhC waveguides. The grating couplers are positioned at each waveguide end, and are intentionally misaligned from the waveguide. This forms a low $Q$ FP cavity between two grating couplers, which helps with the estimation of group indices (Fig. 3). Zoom-in views of the (d) waveguide and (e) grating coupler regions.

We first perform photoluminescence (PL) measurements to optically characterize the fabricated PhC waveguides at room temperature (Fig. 3), leveraging the background emissions of nitrogen-vacancy centers formed during the diamond thin film overgrowth process[37] as internal sources. We couple 532 nm continuous-wave light into a waveguide by focusing it to one waveguide end and collect PL signal generated in the waveguide using the same grating coupler. The collected signal is analyzed by a spectrometer and shown in Fig. 3 (a). We show the PL spectra of fabricated waveguides with three different lattice constants of $a$ = 257, 259, and 261 nm. We observe sharp

FP fringes formed by the light reflection at both grating couplers[34] for both even and odd modes, around 730 nm and 690 nm, respectively. The wavelengths of these modes increase as $a$ increases, as expected. The SiV emission peak is also seen around 737 nm (indicated by the gray dashed line). Fig. 3 (b) displays the enlarged view of the PL spectrum for $a = 261$ nm, confirming that the FP fringes of the even waveguide mode (ranging from 715-740 nm) overlap with the SiV emission peak. The fringe spacing decreases as the peak wavelength increases (Fig. 3(b)), indicating that the $n_g$ increases near the Brillouin zone edge of the waveguide mode due to the reduced group velocity. From the observed FP fringes, we estimate $n_g$ for the fabricated waveguide using the following equation: $n_g = \lambda/2L\Delta\lambda^2$. Here, $L$ is the waveguide length of $51a$ while $\lambda$ and $\Delta\lambda$ are the center wavelength of the FP peak and the wavelength difference between two adjacent FP peaks, respectively. The wavelengths of FP peaks are extracted by fitting the PL spectrum with multiple Lorentzian peak functions. Fig. 3(c) shows the measured $n_g$ overlaid with the simulated values (dash line), with the largest measured value being $n_{g,max} = 73$. For comparison between simulated and experimental $n_g$, we ascribe a wavelength offset to the simulated $n_g$ (Fig.1(c)) to compensate for the wavelength differences caused by fabrication imperfection. The simulated $n_g$ with the offset is in good agreement with the experimental data. It is worth noting that the linewidth of the fringe peaks (corresponding to the quality factor) becomes narrower (higher) as it approaches the slow light region (Fig. 3(b)) (supplementary information), which is consistent with previous observations in PhC waveguides[31,34]. These results indicate the successful demonstration of slow light PhC waveguides in diamond.

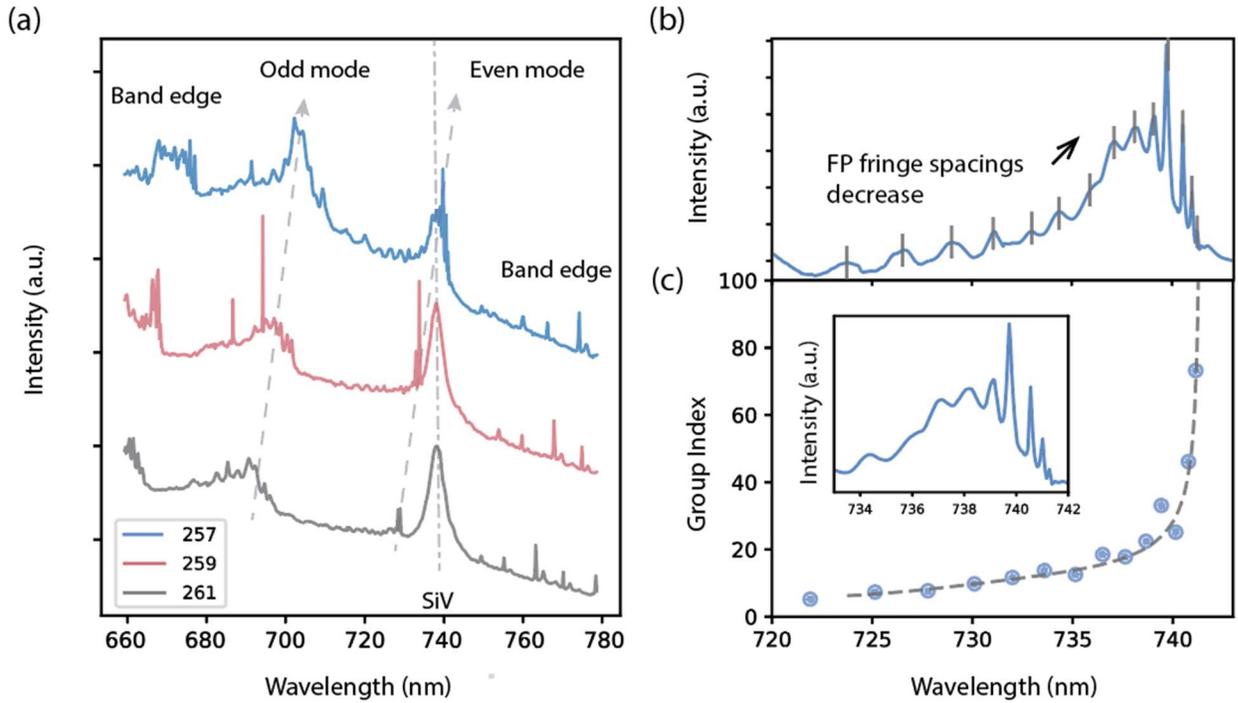

**Fig. 3.** (a) PL spectra measured at room temperature for fabricated waveguides with three different lattice constants ($a$=257, 259, 261 nm). Peaks at 738 nm correspond to the SiV emission (gray

dash line). The gray arrows indicate the positions of odd and even waveguide modes as they shift with the lattice constant. (b) PL spectrum of a waveguide with $a$ = 261 nm shown in (a), after subtracting the broad background fitted by a Gaussian function. (c) Extracted group indices from the spectrum of the FP fringes in (b), obtained by fitting with multiple Lorentz functions. The inset shows the zoom-in view around the slow light region corresponding to (b). The dashed gray curve corresponds to the calculated group indices with a wavelength offset for comparison with the experimental data.

Finally, we perform PL measurements at 5 K (see supplementary information) to investigate the coupling between the PhC waveguide modes and the SiVs (see supplementary information). The SiVs in the waveguide are pumped from above by 520 nm laser light (green arrow in Fig. 4 (a)), and SiV emissions transmitted through the waveguide are collected from one of the grating couplers (red arrow in Fig. 4 (a)) or from free-space. For the optical coupling to SiVs, we use PhC waveguides with $a$ = 261 nm, which has an overlap between the even waveguide modes and the SiV emission, as shown in Fig. 3 (b). Figure 4 (b) shows an energy diagram of the SiVs with four possible optical transitions (A-D). An example of the emission spectrum for a SiV measured from one of the grating couplers is shown in Fig. 4(c), exhibiting the four sharp peaks corresponding to the A-D transitions (see Fig. 4(b)). The observation of SiV emissions via the grating coupler suggests that the SiV is optically coupled to the waveguide mode. We then conduct lifetime measurements on the same SiV using a pulsed laser (supplementary information). The SiV emission is collected at the same spot as the laser excitation spot. We use a spectral filter with a bandwidth of ~50 GHz to filter out only the B or C transition around 737 nm. Based on the measured $n_g$ for the waveguide devices (see supplementary information), we estimate the $n_g$ experienced by the SiVs to be in the range of 11-20. This variation in $n_g$ between different waveguides can be attributed to fabrication imperfections. Figure 4(d) shows the shortest lifetime (1.01 ±0.05 ns, blue, measured at $n_g$ = 11) measured in a waveguide, compared to the longest lifetime (2.32 ±0.12 ns, red) measured in a slab. The decay is fitted with a single exponential with a background offset (black dashed lines). Compared to a typical lifetime of SiV in bulk (~1.7 ns)[21,42], the observed lifetime in a waveguide is reduced by a factor of ~1.7. These observations of the lifetime reduction suggest the spontaneous emission rate of the SiV is enhanced by the Purcell effect in the slow light region. We also statistically investigate the enhancement of the SiV emission coupled to waveguide modes. Fig. 4(e) shows the comparison of lifetime measurements for several SiVs located in two different waveguides with the same structure parameters and on the slab. Overall, the SiV lifetimes in the waveguides (average to 1.30 ns, red star) are shorter than the typical SiV lifetime due to the slow light effect, while those in the slab tend to be the same or longer than bulk SiVs (average to 1.83 ns, blue star). The longer lifetimes in the slab likely are due to the reduced photonic density of states by the slab structure.

The Purcell factor ($F_{ZPL}$) for the investigated ZPL with a lifetime at high $n_g$ region is estimated to be up to ~ 2.2 (see supplementary information), in good agreement with the simulated value of ~ 2.5 (see supplementary information). We also estimate a waveguide-emitter coupling efficiency ($\beta$) of ~72% for the shortest lifetime of 1.0 ns (supplementary information). This value is smaller than values reported in a good cavity-SiV coupled system[21] because of its smaller Purcell enhancement. However, we note that the broadband nature of our devices can allow Purcell enhancement of multiple SiVs without the need for spectral tuning, as is the case in cavity-SiV coupled systems. Our theoretical estimation suggests further optimization, including better spatial matching, could enable $\beta$ ~ 94% with the experimentally achieved $n_g$ ~ 70 (see supplementary information)

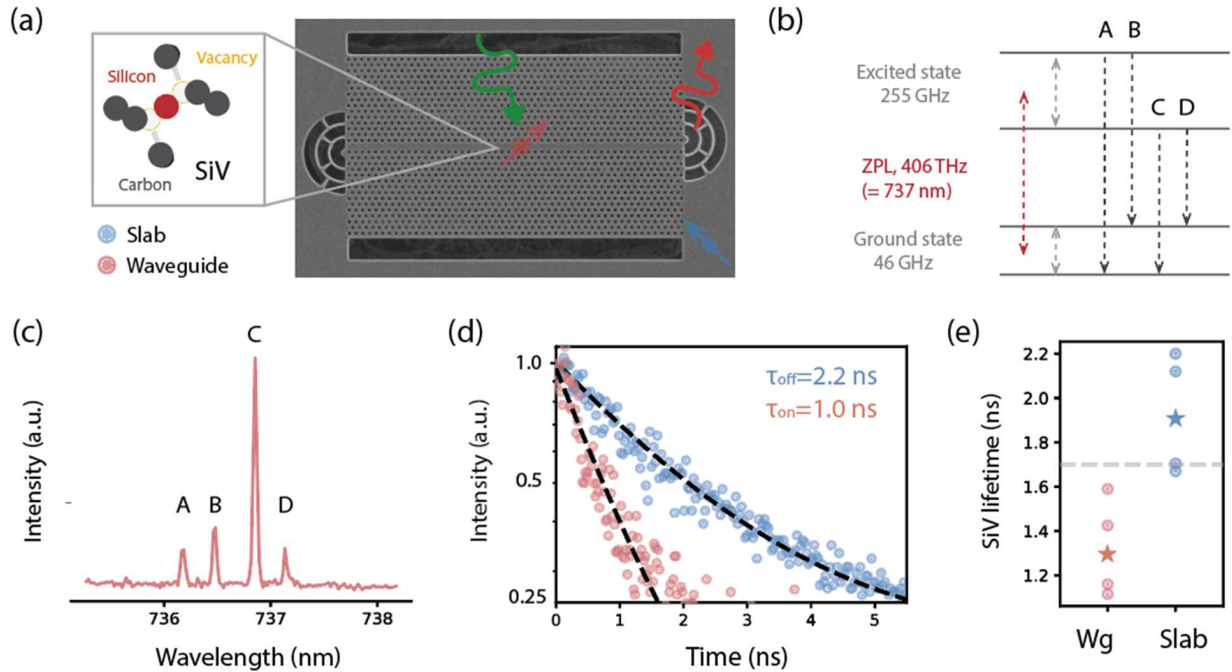

**Fig. 4.** (a) SEM image of a fabricated waveguide. The positions of the excitation and detection are indicated as green arrows and red arrows for measuring the PL spectrum of SiVs. (b) Energy levels of an SiV with 4 possible optical transitions (A~D). (c) PL spectrum of an SiV in the PhC waveguide measured via one of the grating couplers. (d) PL lifetimes of the SiV B and/or C transition in waveguides (red) and in the slab the membrane (blue). The dashed lines correspond to fitting curves with a single exponential function with an offset. (e) Summary of different SiV lifetimes, measured using B and/or C transition: four SiVs are measured in two different waveguides with $a$ = 261 nm and four in the diamond slab. The round dots are the measured data points, and the star dots are the average values of all the waveguide and slab SiV lifetime. The dashed line is the nominal bulk SiV lifetime of 1.7 ns at 4 K.

In summary, we demonstrate slow light 2D PhC waveguides in diamond and study their effect on the emission properties of SiVs. The waveguide modes show a high $n_g$ up to ~70 in the slow light regime. We observe Purcell-enhanced emissions from SiVs in slow-light PhC waveguides, with shorter SiV lifetimes compared to typical bulk ones. Our slow light PhC waveguide approach can be applied to color centers in not only diamond but also other quantum photonic materials, such as silicon carbide[43] and silicon[44]. The proposed fabrication method and PhC waveguide design can also be used to realize more advanced diamond photonic structures, such as chiral waveguide interfaces and topological devices[35], in combination with quantum emitters. Our demonstrations pave the road towards developing novel and efficient photonic hardware for quantum networks based on the color center in diamond.

**Acknowledgments**

We would like to thank Prof. S. Iwamoto for the technical support. This work was supported by AFOSR (Grant No. FA9550-19-1-0376, and FA9550-20-1-0105), ARO MURI (Grant No. W911NF1810432), NSF RAISE TAQS (Grant No. ECCS-1838976), NSF STC (Grant No. DMR-1231319), NSF ERC (Grant No. EEC-1941583), DOE (Grant No. DE-SC0020376), DFG SFB 1375 "NOA" project C5, and ONR (Grant No. N00014-20-1-2425), a research grant from The Mazda Foundation. This work was performed in part at the Center for Nanoscale Systems (CNS), Harvard University. The low-jitter SNSPDs were fabricated at the Jet Propulsion Laboratory, California Institute of Technology, under a contract with the National Aeronautics and Space Administration. Diamond membrane synthesis is based upon work primarily supported by the U.S. Department of Energy Office of Science National Quantum Information Science Research Centers as part of the Q-NEXT center.

# Supplementary Information:

# Purcell enhanced emissions from diamond color centers in slow light photonic crystal waveguides

1. **Optical measurement setup**

For the room-temperature PL measurement of the waveguide modes, we used a commercial spectrometer system (Horiba LabRam Evolution) with free-space off-resonance excitation and collection. The sample is excited with a continuous wave green diode laser (523 nm). The grating selected is 600 gr/mm.

For the low-temperature SiV measurements, we used a confocal setup, as shown in the simplified diagram Fig. S1. The supercontinuum pulsed laser is SuperK EXTREME with a repetition rate of 78 MHz. The 4 K cryo system is Attodry800, and the operating temperature is 5 K. We placed a short pass filter of 700 nm in front of the laser for the off-resonant excitation of the SiVs. A low-jitter NbN SNSPD with impedance-matching tapers and differential readout electronics is provided by our collaborators at JPL[1]. The detectors are optimized for a peak efficiency at 775 nm, but the efficiency is still reasonably high at 737 nm. The spectrometer used is SpectraPro HRS-750, and the grating selected for the SiV spectra is 1800 gr/mm.

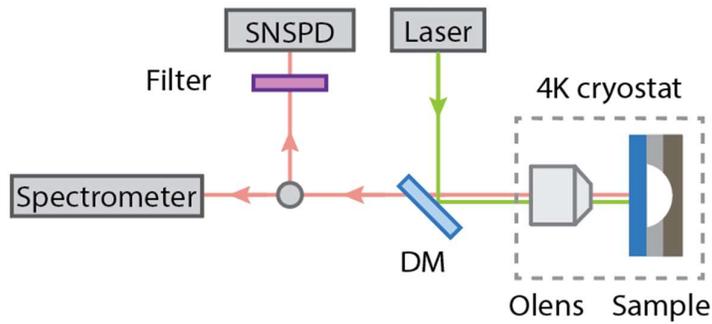

Fig. S1 Schematics of the SiV optical measurement setup. SNSPD: superconducting nanowire single-photon detector; DM: dichroic mirror, reflecting green and passing red; Olens: objective lens.

2. **Measured $Q$ factors (linewidth) of Fabry-Pérot peaks**

Fig. S2 shows measured $Q$ factors (linewidth) extracted from the observed Fabry-Pérot (FP) fringes shown in Fig. 3(b) in the main text. We fitted the FP peaks with multiple Lorentzian peak functions.

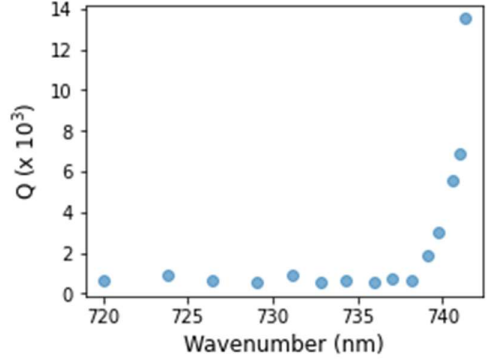

Fig. S2 $Q$ factors for the FP resonance. The data points are extracted from the fitting of FP fringes in Fig. 3(b) in the main text.

3. **Estimation of $n_g$ corresponding to target SiV transition lines**

Fig. S3 (a) and (b) show measured $n_g$ for two waveguide devices with $a$ = 261 nm (named waveguide 1 and 2) which were used for the SiV lifetime experiments in the main text. The theoretical curve is overlaid with the experimental data for better comparison. From the theoretical curve, we can estimate the corresponding $n_g$ at the SiV transition lines (~737nm) to be ~11 and ~20 for waveguide 1 and 2, respectively. The slight difference in position of the Brillouin zone edge of the waveguide modes between waveguide 1 and 2 could be due to fabrication imperfection.

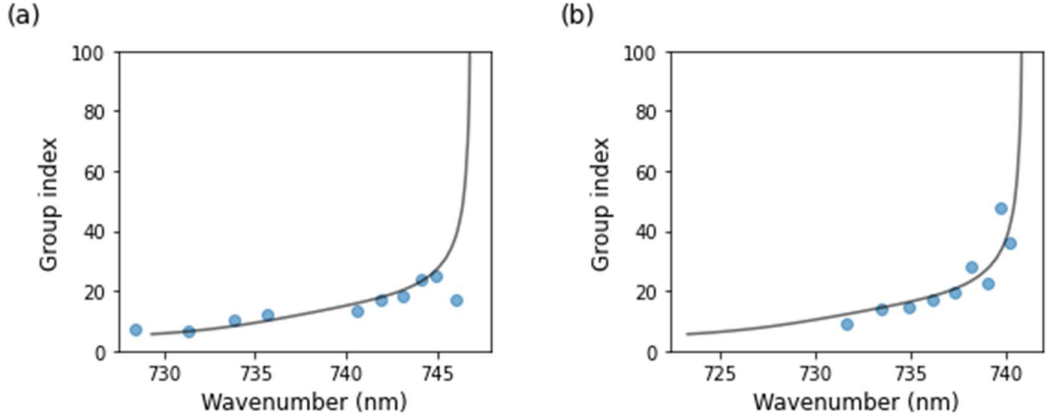

Fig. S3 Measured $n_g$ as a function of wavelength measured for two waveguides with (a) a low $n_g$ (~11) and (b) high $n_g$ (~20) at ~737nm. The dots are experimental data while the gray curve is the theoretical curve with an offset for comparison.

4. **Second-order correlation measurements**

Second-order correlation measurements are performed for a SiV in one of the PhC waveguides using a Hanbury Brown-Twiss setup equipped with two SNSPDs (Fig. S1). We off-resonantly excite the SiV in the PhC waveguide by a pulsed laser (SuperK EXTREME) and collect the PL emission from the C line of the SiV by using a spectral filter with a bandwidth of ~50 GHz. The excitation power is 3.3 mW. Fig. S4 shows the intensity correlation histogram measured at 5K. The second-order correlation function at zero delay time, $g^2(0)$, exhibits an antibunching with a value of 0.47. The non-zero value of $g^2(0)$ could be due to the contribution of the PL emission of waveguide modes. We note that we also confirmed that single SiVs in other thin films that were fabricated using the same technique can exhibit the antibunching feature even in nanostructures [2].

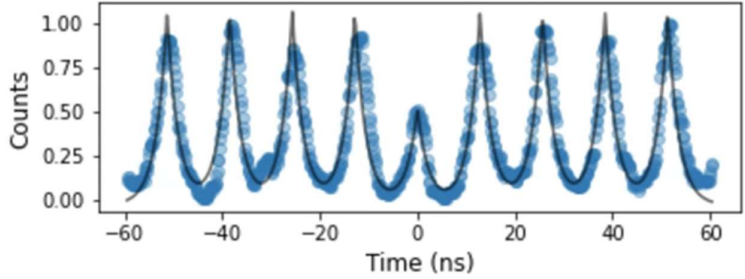

Fig. S4 Second-order correlation measurements. Normalized auto-correlation function $g^2(t)$ measured for a SiV in a PhC waveguide sample. The solid line corresponds to a fitting curve.

5. **Calculation of Purcell factor**

We estimate the Purcell factor of the investigated zero-phonon line (ZPL) $F_{ZPL}$ using the following equation [3]:

$$F_{ZPL} = (\tau_{off}/\tau_{on} - 1)/\xi_{ZPL} \tag{S3}$$

Here $\xi_{ZPL}$ is the fraction of the total emission into the ZPL visible at 4K for an SiV, which is estimated by a product of the Debye–Waller factor of 70%[4] and the branching ratio of 32.5/45.2% into B/C transition at 4K[3]. For $\tau_{off}$, we use the characteristic lifetime of a bulk SiV of 1.7 ns, which is also consistent with the measured SiV lifetime on this sample. For $\tau_{on}$, we use the shortest measured lifetime of a SiV on the resonant waveguide of 1.01 ns.

6. **Calculation of enhancement factor of SiV emission decay rate in waveguides**

We consider a two-dimensional line-defect photonic crystal waveguide in diamond, shown in Fig.1 (a) in the main text (refractive index of the diamond slab $n = 2.4$). The waveguide supports transverse electric (TE) waveguide mode with a group index $n_g$. Under dipole approximation, the SiVs are assumed as point dipoles. The SiV emission decay rate into a waveguide mode $\Gamma_{wg}$ normalized to the decay rate in bulk $\Gamma_0$ is calculated using the following equation [5]:

$$\frac{\Gamma_{wg}}{\Gamma_0} = \frac{3}{4\pi} \frac{(\lambda/n)^2}{S_{eff}} \frac{n_g}{n} \frac{|E(r)\cdot d|^2}{|d|^2|E_{max}|^2} \tag{S2}$$

Here, $\lambda$ and $d$ are the vacuum wavelength and electric dipole moment of the SiV, respectively. $S_{eff}$ is the effective mode defined as:

$$S_{eff} = \iint n(r)^2 |E(r)|^2 d^2r / \max [n(r)^2 |E(r)|^2] \tag{S3}$$

The local electric field at a position $r$ is defined as $E(r)$. We calculate the electric field of the waveguide modes by a 3D plane wave expansion method. We assume that the SiV is primarily coupled to the electric field of $E_y$ to get a larger enhancement factor compared to $E_x$. To calculate the maximum achievable enhancement factor, we take into account a ~30% reduction due to the shallow implantation of the SiVs (~ 40 nm from the top surface, with an ion implantation dose of ~$10^8$) from the maximum electric field at the middle of the diamond slab (80 nm). We also consider the mismatch of the polarization between the local electric field (110) and SiV (111), resulting in an additional reduction of the factor by ~34%.

### 7. Calculation of $\beta$ factor

We roughly estimate the $\beta$ factor using the following equation: $\beta = \gamma_{wg} / (\gamma_{wg} + \gamma_{PhC})$, where $\gamma_{wg}$ and $\gamma_{PhC}$ are the emission rate of SiVs in a waveguide and in photonic crystals (PhCs). We use the measured emission rate of the SiV in a waveguide (~0.83 ns$^{-1}$) for $\gamma_{wg}$. For $\gamma_{PhC}$, we use the previously reported value of ~0.38 ns$^{-1}$ in 2D PhC[6]. In this calculation, we do not include the emission decay into the phonon sideband and non-radiative transitions.

### 8. Estimation of achievable $\beta$ factor

Fig. S5 shows the calculated $n_g$ (top), $Fp$ (middle), and $\beta$ (bottom) for electric field $E_y$, considering the polarization mismatch between a SiV dipole and the local electric field. We assume that the SiV is coupled to the maximum electric field. For the calculation of $\beta$, we use the equation of $\beta$ in section 7, using $\gamma_{wg} = Fp \, \gamma_{bulk}$ (bulk rate $\gamma_{bulk} = 0.59$ ns$^{-1}$) and $\gamma_{PhC} \sim 0.38$ ns$^{-1}$. For the experimentally achievable value of $n_g = 70$ shown in Fig. 3 in the main text, we can estimate the achievable $Fp = 9.4$ and $\beta = 94.3\%$, respectively.

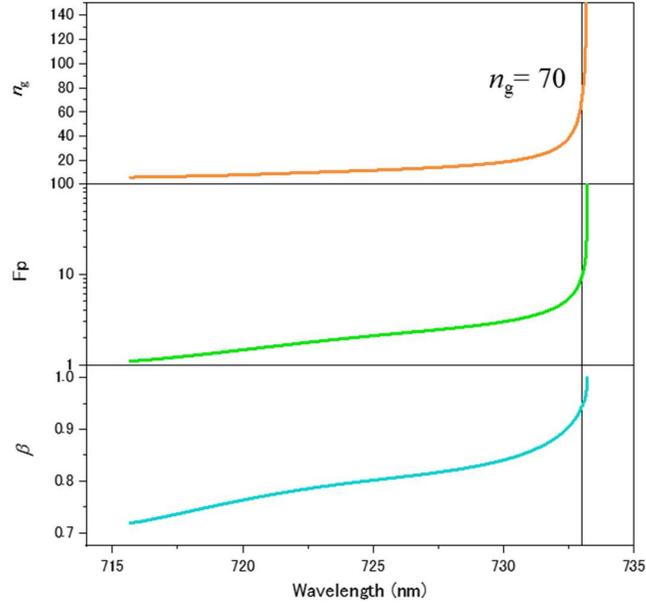

Fig. S5 Calculated $n_g$ (top), $F_p$ (middle), and $β$ (bottom) when the SiV is coupled to the electric field of $E_y$. The vertical line represents the wavelength for $n_g$ =70, corresponding to $F_p$= 9.4 and $β$ = 0.943.